\documentstyle[preprint,eqsecnum,aps]{revtex}
\begin{document}
\draft
\preprint{}
\newcommand{\be}{\begin{equation}}
\newcommand{\ee}{\end{equation}}
\newcommand{\bea}{\begin{eqnarray}}
\newcommand{\eea}{\end{eqnarray}}
\title{Zeta-Functions for Non-Minimal Operators}
\author{
H. T. Cho \footnote[1]{bsp15@twntku10.tku.edu.tw }}
\address{Tamkang University, Department of Physics,\\
Tamsui, Taipei, TAIWAN R.O.C.}
\vskip -.5 truein
\author{
R. Kantowski \footnote[2]{kantowski@phyast.nhn.uoknor.edu}}
\address{
University of Oklahoma, Department of Physics and Astronomy,\\
Norman, OK 73019, USA
}
\maketitle
\begin{abstract}
We evaluate zeta-functions $\zeta(s)$ at $s=0$ for
invariant non-minimal 2nd-order
vector and tensor operators defined on maximally symmetric
even dimensional spaces.
We decompose the operators into their irreducible parts and
obtain their corresponding eigenvalues.
Using these eigenvalues, we
are able to explicitly calculate $\zeta(0)$ for the cases
of Euclidean spaces and $N$-spheres. In the $N$-sphere
case, we make use of the Euler-Maclaurin formula to develop
asymptotic expansions for the required sums. The
resulting $\zeta(0)$ values  for dimensions 2 to 10
are given in the Appendix.
\end{abstract}

\section{Introduction}

The effective potential formalism has been used by
Appelquist and Chodos \cite{AC} and many others
(see \cite{BOS} for a complete list) to consider
the problem of spontaneous
compactification in Kaluza-Klein(KK) theories. The hope was to
explain the smallness of the extra dimensions by using
quantum gravity effects. It was soon realized that the standard
effective action produced results that were dependent on the
choice of the quantum gauge fixing condition
\cite{R-DS,KL} and that all conclusion about stability drawn
from this standard effective action (sometimes called the naive
effective action) were questionable. This problem was
resolved \cite{BO,HKLT}
by the use of a new
effective action, first introduced  by Vilkovisky \cite{GV} and
DeWitt \cite{BD}. This new effective action, now known as the
Vilkovisky-DeWitt(VD) effective action, has the merit of being
gauge choice independent.

However, progress in compactification has never recuperated from
the setback
$\cite{BLO,BKLM}$. The primary reason is that even at the one-loop
level the VD effective
action involves determinants of operators with complicated
non-local terms (in most gauges). In \cite{CK}, we  considered the
six dimensional case for a general background spacetime using
the method of Barvinsky and Vilkovisky \cite{BV} to deal with
the non-local terms. Due to the complexity of this calculation,
it seems quite impossible to push this method to higher dimensions.

The situation can be improved if one chooses the Landau-DeWitt
gauge. Since the VD effective action is independent of gauge
choice, one can of course choose whatever gauge is convenient to
work with, without altering the final result. In the
Landau-DeWitt gauge the non-local terms are identically zero
\cite{FT}.
Although the operators simplify tremendously, they
remain non-minimal (see \cite{BV} and \cite{LOT}),
that is, they involve 2nd-order covariant
derivative terms other than just the Laplacian. More explicitly,
one has to deal with vector operators of the form,
\begin{equation}
{{M_V}^{\alpha}}_{\beta}=-\delta^\alpha_\beta
\Box +a\nabla^\alpha\nabla_\beta-{P^\alpha}_\beta\ ,
\label{MV}
\end{equation}
and the tensor operators of the form,
\begin{equation}
{{M_T}^{\alpha\beta}}_{\rho\sigma}=-\delta^{(\alpha}_{\rho}
\delta^{\beta)}_{\sigma}\Box
+a_{1}\delta^{(\alpha}_{(\rho}\nabla^{\beta)}\nabla_{\sigma)}
+a_{2}g_{\rho\sigma}\nabla^{(\alpha}\nabla^{\beta)}
+a_{3}g^{\alpha\beta}\nabla_{(\rho}\nabla_{\sigma)}
-{P^{\alpha\beta}}_{\rho\sigma}\ ,
\end{equation}
where $A^{(\alpha}B^{\beta)}\equiv (A^{\alpha}B^{\beta}+
A^{\beta}B^{\alpha})/2$. Here we have included the most
frequently encountered
non-minimal second-order terms.

To date progress in evaluating determinants of such operators for
dimensions beyond six has only been made for KK backgrounds of the
form $R_n\times T_{d-n}$ (with $R_n$ usually taken to be flat) and
consequently progress with KK compactification beyond these simple
backgrounds has halted \cite{BO1}.
More interesting backgrounds like $S_d$ or $R_4\times S_{d-4}$ have,
so far, resisted all
efforts \cite{BKO}. In this paper we make progress in evaluating
determinants
of such non-minimal operators on non-flat backgrounds of the form
$S_d$.

Using $\zeta$-function regularization
the one-loop quantum contribution to the
effective action
can be expressed in terms of the
$\zeta$-function
for the appropriate operator as
\begin{equation}
\Gamma_{1}=-{1\over2}\zeta^{'}(0)-\zeta(0)\ {\rm ln}\mu\ ,
\end{equation}
where $\mu$ is the renormalization scale. As a first step
towards obtaining the VD effective action, we shall in this
paper concentrate on the evaluation of the zeta-function
$\zeta(s)$ at $s=0$ for the non-minimal vector and tensor
operators in even dimensions. Of course, these considerations will
also be useful
in the usual effective action formalism when one chooses to work
with gauges other than the Feynman gauge where the
operators are minimal.

In the next section, we obtain eigenvalues for the vector
operator $M_V$  and the tensor operator
$M_T$ in maximally symmetric
spaces, by decomposing the eigenfunctions for
$M_V$ into transverse and longitudinal
parts, and for $M_T$ into
transverse-traceless ($TT$),
longitudinal-transverse-traceless ($LTT$),
longitudinal-longitudinal-traceless ($LLT$)
and trace ($TR$) parts \cite{RO}.
In Section III, we explicitly evaluate $\zeta(0)$
in Euclidean space. Then in Section IV, we extend the results
to N-spheres using the Euler-Maclaurin formula to develop
asymptotic expansions for the relevant summations \cite{BKM}.
Finally, the conclusions are given in Section V. In the
Appendix we summarize
$\zeta(0)$ values  for various cases.

\section{Eigenvalues in Maximally Symmetric Spaces}

The $\zeta$-function for an operator $M$ is defined as
\begin{equation}
\zeta_{M}(s)\equiv \sum_{\lambda}\lambda^{-s}\ ,
\end{equation}
where $\lambda$'s are the eigenvalues of the operator $M$.
Therefore, to calculate $\zeta(s)$ for $M_V$ and $M_T$,
we must first obtain  eigenvalues for these operators. Here
we assume that our background spacetime is a maximally
symmetric space in which the Riemann tensor, the Ricci
tensor, and the scalar curvature are given by,
\begin{eqnarray}
R_{\mu\nu\alpha\beta}&=&\kappa(g_{\mu\alpha}g_{\nu\beta}-
g_{\mu\beta}g_{\nu\alpha})\label{Rie}\ ,\\
R_{\mu\nu}&=&\kappa(N-1)g_{\mu\nu}\label{Rsc}\ ,\\
R&=&\kappa N(N-1)\ ,
\end{eqnarray}
where $N$ is the dimension of the space
and $\kappa$ is a constant.

\subsection{Vector Case}

We first consider the vector operator $M_V$ of (\ref{MV}). For
$M_V$ to be
invariant in maximally symmetric spaces, the function
${P^{\alpha}}_{\beta}$
must be proportional to $\delta^{\alpha}_{\beta}$, i.e.,
\begin{equation}
{{M_V}^{\alpha}}_{\beta}=\delta^{\alpha}_{\beta}
(-\Box-P)+a\nabla^{\alpha}\nabla_{\beta}\ ,
\end{equation}
with $P$ and $a$ constants.
The eigenfunctions $V^{\alpha}$ of $M_V$ can be
decomposed \cite{RO} into a transverse part $T^{\alpha}$,
\begin{equation}
\nabla_{\alpha}T^{\alpha}=0\ ,
\end{equation}
and a longitudinal part $L^{\alpha}$, which is the
gradient of a scalar function $S$,
\begin{equation}
L^{\alpha}=\nabla^{\alpha}S\ ,
\end{equation}
with
\begin{equation}
V^{\alpha}=T^{\alpha}+L^{\alpha}\ .
\end{equation}

Acting with the operator $M_V$ on $T^{\alpha}$ gives
\begin{eqnarray}
{{M_V}^{\alpha}}_{\beta}T^{\beta}&=&
\left[\delta^{\alpha}_{\beta}(-\Box-P)
+a\nabla^{\alpha}\nabla_{\beta}\right]T^{\beta}\ ,\nonumber\\
&=&(-\Box-P)T^{\alpha}\ ,
\label{eigvt}
\end{eqnarray}
because of the transverse property of $T^{\alpha}$. For
$L^{\alpha}$ the result is
\begin{eqnarray}
{{M_V}^{\alpha}}_{\beta}L^{\beta}&=&
\left[\delta^{\alpha}_{\beta}(-\Box-P)
+a\nabla^{\alpha}\nabla_{\beta}\right]\nabla^{\beta}S\ ,\nonumber\\
&=&\left[-(1-a)\Box-P\right]\nabla^{\alpha}S+a[\nabla^{\alpha},
\Box]S\ .
\label{mvl}
\end{eqnarray}
To evaluate the commutator, we use the defining identity
for the Riemann tensor,
\begin{equation}
[\nabla_{\alpha},\nabla_{\beta}]T^{\rho\sigma\cdots}
=-{R^{\xi\rho}}_{\alpha\beta}{T_{\xi}}^{\sigma\cdots}-
{R^{\xi\sigma}}_{\alpha\beta}{{T^{\rho}}_{\xi}}^{\cdots}
+\cdots\ ,
\end{equation}
which for maximally symmetric spaces simplifies to
\begin{equation}
[\nabla_{\alpha},\nabla_{\beta}]T^{\rho\sigma\cdots}
=\kappa(\delta^{\rho}_{\alpha}{T_{\beta}}^{\sigma\cdots}
-\delta^{\rho}_{\beta}{T_{\alpha}}^{\sigma\cdots})+
\kappa(\delta^{\sigma}_{\alpha}
{{T^{\rho}}_{\beta}}^{\cdots}-
\delta^{\sigma}_{\beta}{{T^{\rho}}_{\alpha}}^{\cdots})
+\cdots\ .
\label{ident}
\end{equation}
The commutator in Eq.~(\ref{mvl}) becomes,
\begin{eqnarray}
[\nabla^{\alpha},\Box]S&=&g^{\rho\sigma}[\nabla^{\alpha},
\nabla_{\rho}\nabla_{\sigma}]S\ ,\nonumber\\
&=&\kappa(1-N)\nabla^{\alpha}S\ ,
\end{eqnarray}
and thus,
\begin{equation}
{{M_V}^{\alpha}}_{\beta}L^{\beta}=
\left[-(1-a)\Box-P-a\kappa(N-1)\right]L^{\alpha}\ .
\label{eigvl}
\end{equation}
Using Eqs.~(\ref{eigvt}) and (\ref{eigvl})
one can obtain the eigenvalues of the vector operator $M_V$
when the corresponding
eigenvalues for the Laplacian are known.

\subsection{Tensor Case}

Similar consideration can be applied to the invariant tensor operator
$M_T$. In maximally symmetric spaces, the functions
${P^{\alpha\beta}}_{\rho\sigma}$ can involve only two different
invariant tensors,
$\delta^{(\alpha}_{\rho}\delta^{\beta)}_{\sigma}$ and
$g^{\alpha\beta}g_{\rho\sigma}$.
One can thus write
\begin{equation}
{P^{\alpha\beta}}_{\rho\sigma}=
P\delta^{(\alpha}_{\rho}\delta^{\beta)}_{\sigma}+
Q g^{\alpha\beta}g_{\rho\sigma}\ ,
\end{equation}
with $P$ and $Q$ being constants.
The tensor operator thus becomes,
\begin{equation}
{{M_T}^{\alpha\beta}}_{\rho\sigma}=
(-\Box-P)\delta^{(\alpha}_{\rho}\delta^{\beta)}_{\sigma}
-Q g^{\alpha\beta}g_{\rho\sigma}
+a_{1}\delta^{(\alpha}_{(\rho}\nabla^{\beta)}\nabla_{\sigma)}
+a_{2}g_{\rho\sigma}\nabla^{(\alpha}\nabla^{\beta)}
+a_{3}g^{\alpha\beta}\nabla_{(\rho}\nabla_{\sigma)}\ .
\end{equation}
The eigenfunctions $H^{\alpha\beta}$ of $M_T$ can be
decomposed \cite{RO} into:
\newline
the $TT$ part,
$T^{\alpha\beta}$, where
\begin{eqnarray}
\nabla_{\alpha}T^{\alpha\beta}&=&0\ ,\\
{T_{\alpha}}^{\alpha}&=&0\ ;
\end{eqnarray}
the $LTT$ part,
${L^{T}}^{\alpha\beta}$, where
\begin{equation}
{L^{T}}^{\alpha\beta}=\nabla^{\alpha}T^{\beta}
+\nabla^{\beta}T^{\alpha}\ ,
\end{equation}
for some transverse vector $T^{\alpha}$ with
\begin{equation}
\nabla_{\alpha}T^{\alpha}=0\ ;
\end{equation}
the $LLT$ part,
${L^{L}}^{\alpha\beta}$, where
\begin{equation}
{L^{L}}^{\alpha\beta}=\nabla^{\alpha}\nabla^{\beta}L
+\nabla^{\beta}\nabla^{\alpha}L
-\frac{2}{N}g^{\alpha\beta}\Box L\ ,
\end{equation}
for some scalar function $L$; and the $TR$ part
$g^{\alpha\beta}{H^{\mu}}_{\mu}/N$. Therefore,
\begin{equation}
H^{\alpha\beta}=T^{\alpha\beta}+{L^{T}}^{\alpha\beta}
+{L^{L}}^{\alpha\beta}+\frac{1}{N}g^{\alpha\beta}
{H^{\mu}}_{\mu}\ .
\end{equation}

Acting with $M_T$ on $T^{\alpha\beta}$ and ${L^{T}}^{\alpha\beta}$,
and with the help of the identity in Eq.~(\ref{ident}), we have
\begin{eqnarray}
{{M_{T}}^{\alpha\beta}}_{\rho\sigma}T^{\rho\sigma}&=&
(-\Box-P)T^{\alpha\beta}\ ,
\label{eigtt}\\
{{M_T}^{\alpha\beta}}_{\rho\sigma}{L^{T}}^{\rho\sigma}&=&
\left[-(1-\frac{1}{2}a_{1})\Box-P-a_{1}\kappa\right]
{L^{T}}^{\alpha\beta}\ .
\label{eigltt}
\end{eqnarray}
However, when acting on the $LLT$ part and the $TR$ part,
\begin{eqnarray}
{{M_{T}}^{\alpha\beta}}_{\rho\sigma}
{L^{L}}^{\rho\sigma}
&=&\left[-\left(1-\left(1-\frac{1}{N}\right)a_{1}\right)
\Box-P-\kappa a_{1}(N-1)\right]
{L^{L}}^{\alpha\beta}\nonumber\\
&&\mbox{}+\left[2\left(1-\frac{1}{N}\right)\left(a_{1}+a_{3}N\right)
(\Box+\kappa N)\right]
\left(\frac{1}{N}g^{\alpha\beta}\Box L\right)\ ,\nonumber\\
&&
\label{eigllt}\\
{{M_{T}}^{\alpha\beta}}_{\rho\sigma}\left(\frac{1}{N}g^{\rho\sigma}
H^{\mu}_{\mu}\right)
&=&\left[-\left(1-\left(\frac{a_{1}}{N}+a_{2}+a_{3}\right)\right)
\Box-P-NQ\right]
\left(\frac{1}{N}g^{\alpha\beta}H^{\mu}_{\mu}\right)
\nonumber\\
&&\ +\left[\frac{1}{2}\left(\frac{a_{1}}{N}+a_{2}\right)
(\Box-2\kappa N)\right]
\times
\nonumber\\
&&\ \ \ \left(\nabla^{\alpha}\nabla^{\beta}
+\nabla^{\beta}\nabla^{\alpha}-\frac{2}{N}g^{\alpha\beta}
\Box\right)\left(\frac{1}{\Box}{H^{\mu}}_{\mu}\right)\ .
\label{eigt}
\end{eqnarray}
We see that the second term in Eq.~(\ref{eigllt}) involves
$g^{\alpha\beta}\Box L/N$, which is the trace part of
the function $\nabla^{(\alpha}\nabla^{\beta)}L$, while
the second term in Eq.~(\ref{eigt}) involves
$(\nabla^{\alpha}\nabla^{\beta}
+\nabla^{\beta}\nabla^{\alpha}-\frac{2}{N}g^{\alpha\beta}
\Box)\left(\frac{1}{\Box}{H^{\mu}}_{\mu}\right)$, which belongs to the
$LLT$ part. Hence, the functions in the $LLT$ part and the $TR$
part are coupled together as long as the operator is
non-minimal (unless $a_{2}=a_{3}=-a_{1}/N$).
To find the
eigenfunctions and the corresponding eigenvalues one must take
the appropriate linear combinations of the functions in these
two parts. In the following sections we shall demonstrate explicitly
how this can be done
for Euclidean spaces and N-spheres.

\section{$\zeta$-functions on Euclidean spaces}

In this section, we calculate the $\zeta$-functions for
the vector and tensor operators in N-dimensional Euclidean
spaces. In this simple case where $\kappa =0$,
the eigenvalues for the Laplacian
are just $-k^{2}$, with the eigenfunctions Fourier transformed
to momentum space.

\subsection{Vector Case}

For the vector operator, from
Eqs.~(\ref{eigvt}) and (\ref{eigvl}),
there are $(N-1)$ eigenfunctions in the
transverse part with eigenvalues,
\begin{equation}
\lambda_{T}=k^{2}-P\ ,
\end{equation}
and one eigenfunction in the longitudinal part with eigenvalue,
\begin{equation}
\lambda_{L}=(1-a)k^{2}-P\ .
\end{equation}
Thus the $\zeta$-function is,
\begin{equation}
\zeta^{V}_{N}(s)=(N-1)\sum_{k}(k^2-P)^{-s}+
\sum_{k}[(1-a)k^2-P]^{-s}\ .
\end{equation}
The sum over $k$ is an integral because $k$ is a continuous
variable,
\begin{eqnarray}
\sum_{k}(k^2-P)^{-s}&=&
V_{N}\int\frac{d^{N}\! k}{(2\pi)^{N}}(k^2-P)^{-s}\ ,\nonumber\\
&=&\frac{V_{N}}{\Gamma(s)}\int^{\infty}_{0}d\tau{\tau}^{s-1}
\int\frac{d^{N}\! k}{(2\pi)^{N}}e^{-\tau(k^2-P)}\ ,\nonumber\\
&=&\frac{(-1)^{\frac{N}{2}-s}V_{N}}{(4\pi)^{N/2}}
\frac{\Gamma\left(s-\frac{N}{2}\right)}{\Gamma(s)}
P^{\frac{N}{2}-s}\ ,
\label{sumk}
\end{eqnarray}
where $V_{N}$ is the volume of the $N$-dimensional space
($N$ = even).
Therefore,
\begin{equation}
\zeta^{V}_{N}(s)=\frac{(-1)^{\frac{N}{2}-s}V_{N}}{(4\pi)^{N/2}}
\frac{\Gamma \left(s-\frac{N}{2}\right)}{\Gamma(s)}
\left[(N-1)+(1-a)^{-N/2}\right]P^{\frac{N}{2}-s}\ ,
\end{equation}
and
\begin{equation}
\zeta^{V}_{N}(0)=\frac{V_{N}}{(4\pi)^{N/2}(N/2)!}
\left[(1-a)^{-N/2}+(N-1)\right]P^{N/2}\ .
\end{equation}

\subsection{Tensor Case}

For the tensor operator, there are $\frac{1}{2}(N-2)(N+1)$
eigenfunctions in the $TT$ part with the eigenvalues,
\begin{equation}
\lambda_{TT}=k^{2}-P\ ,
\end{equation}
and $(N-1)$ eigenfunctions in the $LTT$ part with eigenvalues,
\begin{equation}
\lambda_{LTT}=\left(1-\frac{1}{2}a_{1}\right)k^{2}-P\ .
\end{equation}
The $\zeta$-functions corresponding to these two parts
are,
\begin{eqnarray}
\zeta^{TT}_{N}(0)=
\frac{V_{N}}{(4\pi)^{N/2}\left(N/2\right)!}
P^{N/2}\left[\frac{1}{2}(N-2)(N+1)\right]\ ,\\
\zeta^{LTT}_{N}(0)=
\frac{V_{N}}{(4\pi)^{N/2}\left(N/2\right)!}
P^{N/2}\left[(N-1)\left(\frac{2}{2-a_{1}}\right)^{N/2}\right]\ .
\end{eqnarray}
The functions in the $LLT$ and the $TR$ parts are coupled
together as shown in Section II.
By diagonalizing the matrix of which the elements
are given by the coefficients in
Eqs.~(\ref{eigllt}) and (\ref{eigt}),
we can see that the two eigenvalues
$\lambda_{1}$ and $\lambda_{2}$ corresponding to these two
parts satisfy the equations,
\begin{eqnarray}
\lambda_{1}+\lambda_{2}&=&\alpha_{e}k^2+\gamma_{e}\ ,\\
\lambda_{1}\lambda_{2}&=&A_{e}k^4+C_{e}k^2+E_{e}\ ,
\end{eqnarray}
where
\begin{eqnarray}
\alpha_{e}&=&2-(a_{1}+a_{2}+a_{3})\ ,
\label{ecoi}\\
\gamma_{e}&=&-2P-NQ\ ,\\
A_{e}&=&1-(a_{1}+a_{2}+a_{3})-a_{2}a_{3}(N-1)\ ,
\label{ecoA}\\
C_{e}&=&-[2-(a_{1}+a_{2}+a_{3})]P-[N-(N-1)a_{1}]Q\ ,\\
E_{e}&=&P(P+NQ)\ .
\label{ecof}
\end{eqnarray}
Since $\lambda_{1}$ and $\lambda_{2}$ are not
polynomials in $k^2$, it is very difficult
to do the $k$-integral to obtain the corresponding
$\zeta$-functions. In fact, we have
\begin{eqnarray}
\lambda_{1}&=&\frac{1}{2}\left[(\alpha_{e}k^{2}+\gamma_{e})
+\sqrt{(\alpha_{e}k^{2}+\gamma_{e})^{2}-4(A_{e}k^{4}+
C_{e}k^{2}+E_{e})}\right]\ ,\\
\lambda_{2}&=&\frac{1}{2}\left[(\alpha_{e}k^{2}+\gamma_{e})
-\sqrt{(\alpha_{e}k^{2}+\gamma_{e})^{2}-4(A_{e}k^{4}+
C_{e}k^{2}+E_{e})}\right]\ .
\end{eqnarray}
However, we are interested
in the $\zeta$-function at $s=0$, and this depends only on
the small $\tau$ behavior in the integrand of the $\tau$-integral
like the one in Eq.~(\ref{sumk}).
To extract the small $\tau$ behavior from the integral over $k$,
we need only to concentrate on the part of large $k$.
Hence, we can expand $\lambda_{1}$ and $\lambda_{2}$ as power
series in $1/k^{2}$, and we shall see in the following that only the
first few terms will contribute to
$\zeta^{LLT+TR}_{N}(0)$. Expanding,
\begin{eqnarray}
&&-\frac{1}{2}k^{2}\left[\sqrt{\left(\alpha_{e}
+\frac{\gamma_{e}}{k^{2}}\right)^{2}-4\left(A_{e}+\frac{C_{e}}{k^{2}}
+\frac{E_{e}}{k^{4}}\right)}-\left(\alpha_{e}
+\frac{\gamma_{e}}{k^{2}}\right)
\sqrt{1-\frac{4A_{e}}{\alpha_{e}^{2}}}\right]\nonumber\\
&\equiv&\eta_{0}+\frac{\eta_{1}}{k^2}+\frac{\eta_{2}}{k^{4}}+\cdots\ ,
\label{etaexp}
\end{eqnarray}
we can evaluate the $\zeta$-function
for the $LLT$ and the $TR$ parts,
\begin{eqnarray}
\zeta^{LLT+TR}_{N}(0)
&=&\lim_{s\rightarrow 0}V_{N}\int\frac{d^{N}\! k}{(2\pi)^N}
\left[\lambda_{1}^{-s}+\lambda_{2}^{-s}\right]\ ,
\nonumber\\
&=&\lim_{s\rightarrow 0}\frac{V_{N}}{\Gamma(s)}\int^{\infty}_{0}
d\tau\tau^{s-1}\int\frac{d^{N}\! k}{(2\pi)^N}
\left[e^{-\tau\lambda_{1}}+e^{-\tau\lambda_{2}}\right]\ ,
\nonumber\\
&=&\lim_{s\rightarrow 0}\frac{V_{N}}{\Gamma(s)}
\int^{\infty}_{0}d\tau\tau^{s-1}e^{-\frac{\tau}{2}\gamma_{e}}
\int\frac{d^{N}\! k}{(2\pi)^N}e^{-\frac{\tau}{2}
\alpha_{e}k^2}\times
\nonumber\\
&&\ \ \left[2-\frac{\tau\alpha_{e}}{2A_{e}}\sqrt{\alpha_{e}^2-4A_{e}}
\left(\eta_{0}+\frac{\eta_{1}}{k^2}+\frac{\eta_{2}}{k^4}
+\cdots\right)\right.
\nonumber\\
&&\ \ \ \
\left. +\frac{\tau^2}{2}
\left(\frac{\alpha_{e}^2(\alpha_{e}^2-2A_{e})}{4A_{e}^2}\right)
\left(\eta_{0}^{2}+\frac{2\eta_{0}\eta_{1}}{k^2}+\cdots\right)
+\cdots\right]\ .
\label{efinal}
\end{eqnarray}
Note that the last step involves a rescaling
\begin{equation}
\tau\left[1\pm\sqrt{1-4A_{e}/\alpha_{e}^{2}}\right]
\rightarrow\tau\ .
\end{equation}
In this power series form, the integrations over $k$ and $\tau$
can be performed.
To be explicit, we consider the $N=2$ case.
The $k$-integral in Eq.~(\ref{efinal}) becomes,
\begin{eqnarray}
&&\int\frac{d^{N}\! k}{(2\pi)^N}e^{-\frac{\tau}{2}
\alpha_{e}k^2}
\left(2-\frac{\tau\alpha_{e}}{2A_{e}}
\sqrt{\alpha_{e}^2-4A_{e}}\, \eta_{0}\right)+\cdots
\nonumber\\
&=&\frac{V_{2}}{4\pi}
\left[\frac{4}{\alpha_{e}\tau}-\frac{1}{A_{e}}
\sqrt{\alpha_{e}^2-4A_{e}}\, \eta_{0}\right]+\cdots\ .
\label{kint}
\end{eqnarray}
We have left out the terms which, after the integration
over $\tau$, will vanish when the limit $s\rightarrow 0$ is taken.
{}From Eq.~(\ref{etaexp}),
\begin{equation}
\eta_{0}=\frac{\alpha_{e}C_{e}-2A_{e}\gamma_{e}}
{\alpha_{e}\sqrt{\alpha_{e}^{2}-4A_{e}}}\ .
\end{equation}
Using this result and Eq.~(\ref{kint}), the sum of the
$\zeta$-functions for the $LLT$ and the $TR$ parts for $N=2$ is
\begin{eqnarray}
\zeta^{LLT+TR}_{2}(0)&=&
\lim_{s\rightarrow 0}\frac{V_{2}}{(4\pi)\Gamma(s)}
\int^{\infty}_{0}d\tau\tau^{s-1}e^{-\frac{\tau}{2}\gamma_{e}}
\left[\frac{4}{\alpha_{e}\tau}-
\frac{1}{A_{e}}\sqrt{\alpha_{e}^2-4A_{e}}\ \eta_{0}\right]\ ,
\nonumber\\
&=&\frac{V_{2}}{4\pi}
\left[\left(\frac{\alpha_{e}}{A_{e}}\right)P-
\left(\frac{a_{1}-2}{A_{e}}\right)Q\right]\ .
\end{eqnarray}
Finally the $\zeta$-function for the tensor operator on
a two-dimensional Euclidean space is,
\begin{eqnarray}
\zeta^{T}_{2}(0)&=&\zeta^{TT}_{2}(0)+\zeta^{LTT}_{2}(0)
+\zeta^{LLT+TR}_{2}(0)\ ,
\nonumber\\
&=&-\frac{V_{2}}{4\pi}
\left[\left(\frac{2}{a_{1}-2}-\frac{\alpha_{e}}{A_{e}}\right)P+
\left(\frac{a_{1}-2}{A_{e}}\right)Q\right]\ ,
\end{eqnarray}
where $\alpha_{e}$ and $A_{e}$ are given in
Eqs.~(\ref{ecoi}) and (\ref{ecoA}) with $N=2$.

One can extend this procedure to higher even-dimensions.
However, the number of terms involved increases very quickly
and the answers are too lengthy to be written down in any
simple way. We choose to list results in the Appendix for
only the special case in which $a_{1}=-2a_{2}$,
$a_{3}=0$, and with dimensions up to ten.
This case is of special interest because
the tensor operator with these parametrization
corresponds to the graviton operator in Einstein gravity with
the covariant gauge fixing Lagragian,
\begin{equation}
{\cal L}_{gf}=-\frac{1}{2}\sqrt{-g}(1+a_{2}^{2})
\left(\nabla^{\rho}h^{\mu}_{\rho}-
\frac{1}{2}\nabla^{\mu}h^{\rho}_{\rho}\right)
\left(\nabla^{\sigma}h_{\mu\sigma}-\frac{1}{2}
\nabla_{\mu}h^{\sigma}_{\sigma}\right)\ .
\end{equation}
Here $h_{\mu\nu}$ is the graviton field, and
$a_{2}\rightarrow\infty$ gives the Landau-DeWitt gauge that
we have mentioned in Section I.

\section{$\zeta$-functions on N-spheres}

In this section we extend the considerations of the last section to
N-spheres. We use the eigenvalues and the degeneracies for
the Laplacian given in \cite{RO}. For spheres,
\begin{equation}
\kappa=\frac{1}{r^{2}}\ ,
\end{equation}
where $r$ is the radius
of the sphere.

\subsection{Vector Case}

{}From \cite{RO}, the eigenvalues and the degeneracies for the
Laplacian of the transverse part of the vector operator are,
\begin{eqnarray}
\Lambda^{T}_{l}(N)&=&-\frac{l(l+N-1)-1}{r^{2}}\ ,\\
D^{T}_{l}(N)&=&\frac{l(l+N-1)(2l+N-1)(l+N-3)!}
{(N-2)!(l+1)!}\ ,
\end{eqnarray}
where $l=1,2,3,\cdots$. For the longitudinal part, they are
\begin{eqnarray}
\Lambda^{L}_{l}(N)&=&-\frac{l(l+N-1)-(N-1)}{r^{2}}\ ,\\
D^{L}_{l}(N)&=&\frac{(2l+N-1)(l+N-2)!}{l!(N-1)!}\ ,
\end{eqnarray}
where $l=1,2,3,\cdots$. Putting these into Eqs.~(\ref{eigvt})
and (\ref{eigvl}), we obtain the eigenvalues for the
transverse part of the vector operator,
\begin{equation}
\lambda^{T}_{l}(N)=\frac{l(l+N-1)-1}{r^{2}}-P\ ,
\end{equation}
with degeneracies $D^{T}_{l}(N)$, and for the longitudinal
part the eigenvalues,
\begin{equation}
\lambda^{L}_{l}(N)=\frac{(1-a)[l(l+N-1)-(N-1)]}{r^{2}}
-P-\frac{a(N-1)}{r^{2}}\ ,
\end{equation}
with degeneracies $D^{L}_{l}(N)$. The $\zeta$-function is thus
given by,
\begin{eqnarray}
\zeta^{V}_{N}(s)&=&\sum^{\infty}_{l=1}[D^{T}_{l}(N)
(\lambda^{T}_{l}(N))^{-s}+D^{L}_{l}(N)(\lambda^{L}_{l}(N))^{-s}]\ ,
\nonumber\\
&\equiv&\zeta^{T}_{N}(s)+\zeta^{L}_{N}(s)\ .
\end{eqnarray}
Consider first $\zeta^{T}_{N}(s)$,
\begin{eqnarray}
\zeta^{T}_{N}(s)&=&\sum^{\infty}_{l=1}
D^{T}_{l}(N)[\lambda^{T}_{l}(N)]^{-s}\ ,\nonumber\\
&=&\sum^{\infty}_{l=0}D^{T}_{l+1}(N)[\lambda^{T}_{l+1}(N)]^{-s}\ ,
\nonumber\\
&=&\frac{1}{\Gamma(s)}\int^{\infty}_{0}
d\tau\tau^{s-1}\sum^{\infty}_{l=0}D^{T}_{l+1}(N)
e^{-\tau\lambda^{T}_{l+1}(N)}\ .
\end{eqnarray}
For example, for $N=2$, we have
\begin{eqnarray}
\zeta^{T}_{2}(s)&=&\frac{1}{\Gamma(s)}\int^{\infty}_{0}
d\tau\tau^{s-1}\sum^{\infty}_{l=0}(2l+3)
e^{-\tau[\frac{(l+1)(l+2)-1}{r^{2}}-P]}\ ,\nonumber\\
&=&\frac{r^{2s}}{\Gamma(s)}\int^{\infty}_{0}d\tau
\tau^{s-1}e^{-\tau(-Pr^{2}+1)}\sum^{\infty}_{l=0}
(2l+3)e^{-\tau(l^2+3l)}\ .
\label{zetav2}
\end{eqnarray}
Since we just want to evaluate $\zeta$-functions at
$s=0$, we can concern ourselves with the
small $\tau$ behavior of the integrand above. It is
sufficient to have an asymptotic expansion of the $l$-sum
for small $\tau$ to evaluate Eq.~(\ref{zetav2}).
This can be achieved
using the Euler-Maclaurin formula \cite{BKM},
\begin{equation}
\sum^{\infty}_{l=0}f(l)=\int^{\infty}_{0}dl\ f(l)+\frac{1}{2}
[f(\infty)+f(0)]+\sum^{\infty}_{k=1}\frac{B_{2k}}{(2k)!}
[f^{(2k-1)}(\infty)-f^{(2k-1)}(0)]\ ,
\end{equation}
where $B_{2k}$ are the Bernoulli numbers. Using this formula,
the sum in $\zeta^{T}_{2}(s)$ can be expanded into,
\begin{equation}
\sum^{\infty}_{l=0}(2l+3)e^{-\tau(l^2+3l)}=\frac{1}{\tau}+
\frac{4}{3}+O(\tau)\ .
\end{equation}
When this asymptotic expansion is put back into Eq.~(\ref{zetav2}),
the terms with order $\tau$ or higher in the expansion will
vanish as $s\rightarrow 0$ (because $\Gamma(s)\sim 1/s$).
Therefore,
\begin{eqnarray}
\zeta^{T}_{2}(0)&=&\lim_{s\rightarrow 0}\frac{1}{\Gamma(s)}
\int^{\infty}_{0}d\tau\ \tau^{s-1}e^{-\tau(-Pr^{2}+1)}
\left(\frac{1}{\tau}+\frac{4}{3}\right)\ ,\nonumber\\
&=&(Pr^{2})+\frac{1}{3}\ .
\end{eqnarray}
Similarly, for $\zeta^{L}_{2}(s)$, we have
\begin{equation}
\zeta^{L}_{2}(0)=\left(\frac{1}{1-a}\right)(Pr^{2})+
\left(\frac{1}{1-a}-\frac{2}{3}\right)\ .
\end{equation}
Hence,
\begin{eqnarray}
\zeta^{V}_{2}(0)&=&\zeta^{T}_{2}(0)+\zeta^{L}_{2}(0)\ ,\nonumber\\
&=&\left(\frac{1}{1-a}+1\right)(Pr^{2})
+\left(\frac{1}{1-a}-\frac{1}{3}\right)\ .
\end{eqnarray}
We have extended this procedure up to $N=10$,
and the result is
summarized in the Appendix.

\subsection{Tensor Case}

{}From \cite{RO}, the eigenvalues and degeneracies for the
Laplacian of the $TT$ part of the tensor operator are,
\begin{eqnarray}
\Lambda^{TT}_{l}(N)&=&-\frac{l(l+N-1)-2}{r^{2}}\ ,\nonumber\\
D^{TT}_{l}(N)&=&\frac{(N+1)(N-2)(l+N)(l-1)(2l+N-1)(l+N-3)!}
{2(N-1)!(l+1)!}\ ,
\end{eqnarray}
with $l=2,3,\cdots$. For the $LTT$ part, they are
\begin{eqnarray}
\Lambda^{LTT}_{l}(N)&=&-\frac{l(l+N-1)-(N+2)}{r^{2}}\ ,\nonumber\\
D^{LTT}_{l}(N)&=&\frac{l(l+N-1)(2l+N-1)(l+N-3)!}{(N-2)!(l+1)!}\ ,
\end{eqnarray}
with $l=2,3,\cdots$. For the $LLT$ part, they are
\begin{eqnarray}
\Lambda^{LLT}_{l}(N)&=&
-\frac{l(l+N-1)-2N}{r^2}\ ,
\nonumber\\
D^{LLT}_{l}(N)&=&\frac{(2l+N-1)(l+N-2)!}{l!(N-1)!}\ ,
\end{eqnarray}
with $l=2,3,\cdots$.
For the $TR$ part, they are
\begin{eqnarray}
\Lambda^{TR}_{l}(N)&=&-\frac{l(l+N-1)}{r^{2}}\ ,\nonumber\\
D^{TR}_{l}(N)&=&\frac{(2l+N-1)(l+N-2)!}{l!(N-1)!}\ ,
\end{eqnarray}
with $l=0,1,2,\cdots$. Thus, from Eqs.~(\ref{eigtt}) and
(\ref{eigltt}), the eigenvalues for the $TT$ part of the
tensor operator are,
\begin{equation}
\lambda^{TT}_{l}(N)=\frac{l(l+N-1)-2}{r^{2}}-P\ ,
\end{equation}
with degeneracies $D^{TT}_{l}(N)$, and for the $LTT$ part
the eigenvalues are,
\begin{equation}
\lambda^{LTT}_{l}(N)=
\left(1-\frac{a_{1}}{2}\right)
\left(\frac{l(l+N-1)-(N+2)}{r^{2}}\right)
-P-\frac{a_{1}}{r^{2}}\ ,
\end{equation}
with degeneracies $D_{l}^{LTT}(N)$.
One can see that they are very similar to the ones in the vector
case, so it is straight forward to obtain the $\zeta$-function
corresponding to these two parts, $\zeta^{TT}_{N}(0)$ and
$\zeta^{LLT}_{N}(0)$, of the tensor operator using the same method
as in the case of the vector operator.
For example, for $N=2$ we have
\begin{eqnarray}
\zeta^{TT}_{2}(0)&=&0\ ,\nonumber\\
\zeta^{LLT}_{2}(0)&=&-\frac{2}{(a_{1}-2)}(Pr^{2})
-\left(\frac{4}{a_{1}-2}+\frac{5}{3}\right)\ .
\end{eqnarray}

For the $LLT$ and the $TR$ parts, the situation is more complicated
because the functions are coupled together. As in the Euclidean
case, we can obtain the following relations for the eigenvalues
from Eqs.~(\ref{eigllt}) and (\ref{eigt}),
\begin{eqnarray}
\lambda_{1}+\lambda_{2}&=&
\frac{1}{r^{2}}[\alpha l^{2}+\beta l+\gamma]\ ,\\
\lambda_{1}\lambda_{2}&=&
\frac{1}{r^{4}}(Al^{4}+Bl^{3}+Cl^{2}+Dl+E)\ ,
\end{eqnarray}
where
\begin{eqnarray}
\alpha&=&2-(a_{1}+a_{2}+a_{3})\ ,
\label{c11}\\
\beta&=&(N-1)\alpha\ ,
\label{c12}\\
\gamma&=&-2(Pr^{2})-NQr^{2}+(N-1)(a_{1}-2)-2\ ,
\label{c13}\\
A&=&1-(a_{1}+a_{2}+a_{3})-(N-1)a_{2}a_{3}\ ,
\label{c21}\\
B&=&2(N-1)A\ ,
\label{c22}\\
C&=&\frac{D}{N-1}+(N-1)^{2}A\ ,
\label{c23}\\
D&=&-(N-1)\alpha(Pr^{2})+(N-1)^{2}(a_{1}-2)Qr^{2}
-(N-1)(2-N)Qr^{2}\nonumber\\
&&\mbox{}\mbox{}-N(N-1)A-N(N-1)\alpha
-(N-1)^{2}(a_{1}-2)+(N-1)(2-N)\ ,
\label{c24}\\
E&=&-[\gamma+(Pr^{2}+NQr^{2})][Pr^{2}+NQr^{2}]\ .
\label{c25}
\end{eqnarray}

Note that the degeneracies
$D^{LLT}_{l}(N)$ and $D^{TR}_{l}(N)$ are the same because they
are both concerned with scalar functions, $L$ and
${H^{\mu}}_{\mu}$. However, $l$ starts from 2 in the $LLT$ part,
while $l$ starts from 0 in the $TR$ part. Thus the cases
with $l=0$ and 1 in the $TR$ part have to be separated out.
The $\zeta$-function for these two coupled parts becomes,
\begin{eqnarray}
\zeta^{LLT+TR}_{N}(s)
&=&\sum^{\infty}_{l=2}D^{TR}_{l}(N)\left[\lambda_{1}^{-s}
+\lambda^{-s}_{2}\right]+D^{TR}_{0}(N)(-P-NQ)^{-s}\nonumber\\
&&\ +D^{TR}_{1}(N)
\left[\left(1-\left(\frac{a_{1}}{N}+a_{2}+a_{3}\right)\right)
\frac{N}{r^{2}}-P-NQ\right]^{-s}\ .
\end{eqnarray}
As $s\rightarrow 0$, which is the limit we shall ultimately
take,
\begin{eqnarray}
\zeta^{LLT+TR}_{N}(0)
&=&\lim_{s\rightarrow 0}\sum^{\infty}_{l=2}
D^{TR}_{l}(N)[\lambda^{-s}_{1}+\lambda^{-s}_{2}]
+D^{TR}_{0}(N)+D^{TR}_{1}(N)\ ,\nonumber\\
&=&\lim_{s\rightarrow 0}\sum^{\infty}_{l=0}
D^{TR}_{l}(N)\left[\lambda^{-s}_{1}+
\lambda^{-s}_{2}\right]-(N+2)\ ,
\label{zlltt}
\end{eqnarray}
because
\begin{eqnarray}
D^{LLT}_{0}(N)&=&D^{TR}_{0}(N)=1\ ,\\
D^{LLT}_{1}(N)&=&D^{TR}_{1}(N)=N+1\ .
\end{eqnarray}
For $\lambda_{1}$ and $\lambda_{2}$, we have
\begin{equation}
\lambda_{1}=\frac{1}{2r^{2}}[(\alpha l^{2}+\beta l+\gamma)
+\sqrt{(\alpha l^{2}+\beta l+\gamma)^{2}-4(Al^{4}+
Bl^{3}+Cl^{2}+Dl+E)}]\ ,
\end{equation}
\begin{equation}
\lambda_{2}=\frac{1}{2r^{2}}[(\alpha l^{2}+\beta l+\gamma)
-\sqrt{(\alpha l^{2}+\beta l+\gamma)^{2}-4(Al^{4}+
Bl^{3}+Cl^{2}+Dl+E)}]\ .
\end{equation}
Since the eigenvalues are not polynomials in $l$,
we cannot (as in the vector case) apply the
Euler-Maclaurin formula directly to
obtain the asymptotic series in $\tau$.
As was done in the Euclidean case,
we first expand $\lambda_{1}$ and $\lambda_{2}$ as power
series in $1/l$, and then use the Euler-Maclaurin formula
to evaluate the sums over $l$. Suppose that,
\begin{eqnarray}
-\frac{1}{2}l^{2}\!\!\!\!\!\!\!
&&\left[\sqrt{\left(\alpha+\frac{\beta}{l}
+\frac{\gamma}{l^{2}}\right)^{2}
-4\left(A+\frac{B}{l}+\frac{C}{l^{2}}
+\frac{D}{l^{3}}+\frac{E}{l^{4}}\right)}
-\left(\alpha+\frac{\beta}{l}
+\frac{\gamma}{l^{2}}\right)
\sqrt{1-\frac{4A}{\alpha^{2}}}\right]\nonumber\\
&\equiv&\xi_{0}+\frac{\xi_{1}}{l}+\frac{\xi_{2}}{l^{2}}+\cdots\ .
\label{lexp}
\end{eqnarray}
There is no $l^2$ term in this expansion because we have explicitly
subtracted it out. The $l$ term also vanishes as a result of the form
of $\beta$ and $B$ in Eqs.~(\ref{c12}) and (\ref{c22}). With this
expansion we can evaluate the sum in Eq.~(\ref{zlltt}),
\begin{eqnarray}
&&\lim_{s\rightarrow 0}\sum^{\infty}_{l=0}D^{TR}_{l}(N)
[\lambda^{-s}_{1}+\lambda^{-s}_{2}]\nonumber\\
&=&\lim_{s\rightarrow 0}\sum^{\infty}_{l=0}
D^{TR}_{l}(N)\frac{1}{\Gamma(s)}\int^{\infty}_{0}
d\tau\tau^{s-1}\left[e^{-\tau\lambda_{1}}+
e^{-\tau\lambda_{2}}\right]\ ,
\nonumber\\
&=&\lim_{s\rightarrow 0}\frac{1}{\Gamma(s)}
\int^{\infty}_{0}d\tau\tau^{s-1}e^{-\frac{\tau}{2}\gamma}
\sum^{\infty}_{l=0}D^{TR}_{l}(N)e^{-\frac{\tau}{2}
(\alpha l^2+\beta l)}\times
\nonumber\\
&&\ \ \left[2-\frac{\tau\alpha}{2A}\sqrt{\alpha^2-4A}
\left(\xi_{0}+\frac{\xi_{1}}{l}+\frac{\xi_{2}}{l^2}+\cdots\right)\right.
\nonumber\\
&&\ \ \ \ \left.
+\frac{\tau^2}{2}\left(\frac{\alpha^2(\alpha^2-2A)}{4A^2}\right)
\left(\xi_{0}^{2}+\frac{2\xi_{0}\xi_{1}}{l}+\cdots\right)+\cdots
\right]\ .
\label{final}
\end{eqnarray}
This last step involves a scaling
\begin{equation}
\frac{\tau}{r^{2}}\left[1\pm\sqrt{1-\frac{4A}{\alpha^{2}}}\right]
\rightarrow\tau\ .
\end{equation}
In this form, one can now apply the Euler-Maclaurin formula to
obtain an asymptotic series for small $\tau$ for the sum. To be
explicit, we consider the $N=2$ case, where
the sum in Eq.~(\ref{final}) becomes,
\begin{eqnarray}
&&\sum_{l=0}^{\infty}(2l+1)e^{-\frac{\tau}{2}(\alpha l^2+\beta l)}
\left(2-\frac{\tau\alpha}{2A}\sqrt{\alpha^2-4A}\xi_{0}\right)
+\cdots
\nonumber\\
&=&\frac{4}{\alpha\tau}
+\left(\frac{2}{3}-\frac{1}{A}
\sqrt{\alpha^2-4A}\, \xi_{0}\right)+\cdots\ .
\end{eqnarray}
Again we have left out the terms which, after the integration
over $\tau$, will vanish when the limit $s\rightarrow 0$ is taken.
Putting the expressions from Eq.~(\ref{c11}) to Eq.~(\ref{c25}),
with $N=2$,
into the expansion in Eq.~(\ref{lexp}), we have
\begin{equation}
\xi_{0}=\frac{\alpha D-2A\gamma}{\alpha\sqrt{\alpha^2-4A}}\ .
\end{equation}
Therefore,
\begin{eqnarray}
&&\lim_{s\rightarrow 0}\sum^{\infty}_{l=0}
D^{TR}_{l}(2)\left[\lambda^{-s}_{1}+\lambda^{-s}_{2}
\right]
\nonumber\\
&=&\lim_{s\rightarrow 0}\frac{r^{2s}}{\Gamma(s)}
\int^{\infty}_{0}d\tau\tau^{s-1}e^{-\frac{\tau}{2}\gamma}
\left[\frac{4}{\alpha\tau}+\frac{2}{3}-
\frac{1}{A}\left(\sqrt{\alpha^2-4A}\, \xi_{0}\right)\right]\ ,
\nonumber\\
&=&\left(\frac{\alpha}{A}\right)(Pr^2)-
\left(\frac{a_{1}-2}{A}\right)(Qr^2)+
\left[\frac{2}{3A}(3\alpha+4A)+\frac{a_{1}-2}{A}\right]\ .
\end{eqnarray}
Putting this result into Eq.~(\ref{zlltt}), the
$\zeta$-function for the $LLT$ and the $TR$ parts for $N=2$ is
\begin{equation}
\zeta^{LLT+TR}_{2}(0)=
\left(\frac{\alpha}{A}\right)(Pr^2)-
\left(\frac{a_{1}-2}{A}\right)(Qr^2)+
\left[\frac{2}{3A}(3\alpha-2A)+\frac{a_{1}-2}{A}\right]\ .
\end{equation}
Finally the $\zeta$-function for the tensor operator on 2-sphere
is,
\begin{eqnarray}
\zeta^{T}_{2}(0)&=&\zeta^{TT}_{2}(0)+\zeta^{LTT}_{2}(0)
+\zeta^{LLT+TR}_{2}(0)\ ,
\nonumber\\
&=&-\left(\frac{2}{a_{1}-2}-\frac{\alpha}{A}\right)(Pr^2)-
\left(\frac{a_{1}-2}{A}\right)(Qr^2)
\nonumber\\
&&\ \ +\left[-\frac{4}{a_{1}-2}+\frac{1}{A}(2\alpha-3A)+
\frac{a_{1}-2}{A}\right]\ ,
\end{eqnarray}
where $\alpha$ and $A$ are given in Eqs.~(\ref{c11}) and
(\ref{c21}) with $N=2$.
One can extend this procedure to higher even-dimensions.
The results for $N=2$ through 10
with $a_{1}=-2a_{2}$ and $a_{3}=0$
are listed in the Appendix.

\section{Conclusions}

We have shown how to evaluate the zeta-function at zero, $\zeta(0)$,
for certain non-minimal vector and tensor operators. The procedure is
to first
decompose the vector and tensor functions into their
irreducible parts. For vectors there are the transverse ($T$)
and the longitudinal ($L$) parts. For tensors there are the
$TT$, the $LTT$, the $LLT$ and the $TR$ parts.
Then evaluate the eigenvalues for the various parts
of each operator. Due to the fact that the tensor operator
is non-minimal, the $LLT$ and
the $TR$ parts are in fact coupled together and the
eigenvalues for these two parts are complicated.
However, we have shown that it is still possible to obtain
$\zeta(0)$ by use of
the appropriate series expansion for the eigenvalues.
Using this procedure we explicitly evaluated $\zeta(0)$
for Euclidean spaces and N-spheres for even dimensions up to ten,
and summarized the results in the Appendix. Other techniques have been
developed to successfully deal with flat backgrounds \cite{LOT}; however,
our use of the Euler-Maclaurin formula \cite{BKM} has allowed us to now deal
with
more interesting backgrounds.

Although the above procedure gets more tedious as
one goes to higher dimensions, there is no conceptual
difficulty in doing so. One can extend this method
to dimensions higher than ten, to Kaluza-Klein spacetimes
like $M^{4}\times S^{N}$ \cite{LOT},
and to more general coset spaces for which eigenvalues
of the corresponding Laplacians are known.

The method developed here is general enough to be useful
in many circumstances when one-loop quantum effects are
calculated in gauge theories with general gauge conditions.
What we have in mind in particular is
the calculation of the VD effective potentials in
Kaluza-Klein spaces. We are also interested in
using the VD formalism to evaluate the gauge-independent
trace anomaly for gravitons \cite{CK2}.
The explicit evaluation
of this trace anomaly in different spacetimes
will be possible by making use of the $\zeta$-functions derived
here.

\acknowledgements

H. T. Cho is supported by the National Science Council
of the Republic of China under contract number
NSC 83-0208-M-032-034. R. Kantowski is supported by the
U.S. Department of Energy.

\appendix
\section*{}

In this appendix we summarize the $\zeta$-function values that
we have obtained for the non-minimal vector operator,
\begin{equation}
{{M_V}^{\alpha}}_{\beta}=\delta^\alpha_\beta
(-\Box -P)+a\nabla^\alpha\nabla_\beta
\label{appmv}
\end{equation}
and the non-minimal tensor operator,
\begin{equation}
{{M_T}^{\alpha\beta}}_{\rho\sigma}=\delta^{(\alpha}_{\rho}
\delta^{\beta)}_{\sigma}(-\Box-P)-g^{\alpha\beta}g_{\rho\sigma}Q
-2a_{2}\delta^{(\alpha}_{(\rho}
\nabla^{\beta)}\nabla_{\sigma)}+a_{2}g_{\rho\sigma}
\nabla^{(\alpha}\nabla^{\beta)}
\label{appmt}
\end{equation}

\subsection{Vector Case}

For $N$-dimensional Euclidean spaces, the $\zeta$-function values
for $M_{V}$ in Eq.~(\ref{appmv}) are,
\begin{equation}
\zeta^{V}_{N}(0)=\frac{V_{N}}{(4\pi)^{N/2}(N/2)!}
\left[(1-a)^{-N/2}+(N-1)\right]P^{N/2}
\end{equation}
where $V_{N}$ is the volume of the $N$-dimensional space.

While for $N$-spheres with radius $r$,
where $N=2,4,6,8,10$, we have,
\begin{eqnarray}
\zeta^{V}_{2}(0)&=&\frac{1}{3(1-a)}
\left[3(2-a)(Pr^2)+(2+a)\right]
\\
\zeta^{V}_{4}(0)&=&\frac{1}{180(1-a)^2}\times
\nonumber\\
&&\ \left[15(4-6a+3a^2)(Pr^2)^2+30(8-8a+3a^2)(Pr^2)
\right.
\nonumber\\
&&\ \ \left.
+(172+106a-143a^2)\right]
\\
\zeta^{V}_{6}(0)&=&\frac{1}{2520(1-a)^3}
\times
\nonumber\\
&&\ \left[7(6-15a+15a^2-5a^3)(Pr^2)^3
+105(6-13a+12a^2-4a^3)(Pr^2)^2\right.
\nonumber\\
&&\ \ +21(134-203a+129a^2-35a^3)(Pr^2)
\nonumber\\
&&\ \ \left.
+(3394+213a-5358a^2+2626a^3)\right]
\\
\zeta^{V}_{8}(0)&=&\frac{1}{907200(1-a)^4}
\times
\nonumber\\
&&\ \left[45(8-28a+42a^2-28a^3+7a^4)(Pr^2)^4\right.
\nonumber\\
&&\ \
+420(32-104a+150a^2-100a^3+25a^4)(Pr^2)^3
\nonumber\\
&&\ \ +630(280-796a+1038a^2-668a^3+167a^4)
(Pr^2)^2
\nonumber\\
&&\ \
+180(5200-11064a+9930a^2-4724a^3+1001a^4)(Pr^2)
\nonumber\\
&&\ \ \left.
+(1592968-1081132a-2826102a^2+3532148a^3
-1109837a^4)\right]
\\
\zeta^{V}_{10}(0)&=&\frac{1}{59875200(1-a)^5}
\times
\nonumber\\
&&\ \left[33(10-45a+90a^2-90a^3+45a^4-9a^5)(Pr^2)^5
\right.
\nonumber\\
&&\ \ +495(50-215a+420a^2-420a^3+210a^4
-42a^5)(Pr^2)^4
\nonumber\\
&&\ \ +330(2150-8611a+16028a^2-15810a^3
+7905a^4-1581a^5)(Pr^2)^3
\nonumber\\
&&\ \ +330(29018-103119a+172158a^2-160184a^3
+78570a^4-15714a^5)(Pr^2)^2
\nonumber\\
&&\ \ +1485(40002-112777a+135030a^2-89946a^3
+34405a^4-5985a^5)(Pr^2)
\nonumber\\
&&\ \
+(129517198-192948785a-167760920a^2+502687820a^3
\nonumber\\
&&\ \ \left.
-351902170a^4+82355474a^5)\right]
\end{eqnarray}

\subsection{Tensor Case}

For the $N$-dimensional Euclidean spaces, where $N=2,4,6,8,10$,
the $\zeta$-function values
for $M_{T}$ in Eq.~(\ref{appmt}) are,
\begin{eqnarray}
\zeta^{T}_{2}(0)&=&\frac{V_{2}}{4\pi(1+a_{2})}
\left[(3+a_{2})P+2(1+a_{2})Q\right]
\\
\zeta^{T}_{4}(0)&=&\frac{V_{4}}{(4\pi)^2(1+a_{2})^2}
\times
\nonumber\\
&&\ \left[(5+6a_{2}+3a_{2}^{2})P^2
+2(2+6a_{2}+3a_{2}^{2})PQ\right.
\nonumber\\
&&\ \ \left.
+2(2+3a_{2})^{2}Q^2\right]
\\
\zeta^{T}_{6}(0)&=&\frac{V_{6}}{6(4\pi)^3(1+a_{2})^3}\times
\nonumber\\
&&\ \left[3(7+15a_{2}+15a_{2}^{2}+5a_{2}^{3})P^3
+6(3+15a_{2}+15a_{2}^{2}+5a_{2}^{3})P^{2}Q\right.
\nonumber\\
&&\ \ \left.
+12(9+45a_{2}+65a_{2}^{2}+25a_{2}^{3})PQ^2
+8(3+5a_{2})^{3}Q^{3}\right]
\\
\zeta^{T}_{8}(0)&=&\frac{V_{8}}{6(4\pi)^4(1+a_{2})^4}\times
\nonumber\\
&&\ \left[(9+28a_{2}+42a_{2}^{2}+28a_{2}^{3}+7a_{2}^{4})
P^4\right.
\nonumber\\
&&\ \ +2(4+28a_{2}+42a_{2}^{2}+28a_{2}^{3}+7a_{2}^{4})
P^{3}Q
\nonumber\\
&&\ \ +6(16+112a_{2}+238a_{2}^{2}+182a_{2}^{3}+49a_{2}^{4})
P^{2}Q^{2}
\nonumber\\
&&\ \ +8(4+7a_{2})^2(4+14a_{2}+7a_{2}^{2})PQ^{3}
\nonumber\\
&&\ \ \left.
+4(4+7a_{2})^{4}Q^{4}\right]
\\
\zeta^{T}_{10}(0)&=&\frac{V_{10}}{120(4\pi)^5(1+a_{2})^5}\times
\nonumber\\
&&\ \left[5(11+45a_{2}+90a_{2}^{2}+90a_{2}^{3}+45a_{2}^{4}
+9a_{2}^{5})P^5\right.
\nonumber\\
&&\ \ +10(5+45a_{2}+90a_{2}^{2}+90a_{2}^{3}+45a_{2}^{4}
+9a_{2}^{5})P^{4}Q
\nonumber\\
&&\ \ +40(25+225a_{2}+630a_{2}^{2}+720a_{2}^{3}+387a_{2}^{4}
+81a_{2}^{5})P^{3}Q^{2}
\nonumber\\
&&\ \ +80(125+1125a_{2}+3600a_{2}^{2}+5094a_{2}^{3}
+3159a_{2}^{4}+729a_{2}^{5})P^{2}Q^{3}
\nonumber\\
&&\ \ +80(5+9a_{2})^3(5+18a_{2}+9a_{2}^{2})PQ^{4}
\nonumber\\
&&\ \ \left.
+32(5+9a_{2})^{5}Q^{5}\right]
\end{eqnarray}
While for the $N$-spheres, where again $N=2,4,6,8,10$,
we have,
\begin{eqnarray}
\zeta^{T}_{2}(0)&=&\frac{1}{1+a_{2}}
\left[(3+a_{2})Pr^{2}+2(1+a_{2})Qr^{2}+(1-3a_{2})\right]
\\
\zeta^{T}_{4}(0)&=&\frac{1}{18(1+a_{2})^2}\times
\nonumber\\
&&\ \left[3(5+6a_{2}+3a_{2}^{2})(Pr^2)^2
+6(2+6a_{2}+3a_{2}^{2})(Pr^2)(Qr^2)\right.
\nonumber\\
&&\ \ +6(2+3a_{2})^2(Qr^2)^2
+12(5-a_{2}-3a_{2}^{2})(Pr^2)
\nonumber\\
&&\ \ \left.+12(2+5a_{2})(Qr^2)
+2(11-122a_{2}-97a_{2}^{2})\right]
\\
\zeta^{T}_{6}(0)&=&\frac{1}{360(1+a_{2})^3}\times
\nonumber\\
&&\ \left[3(7+15a_{2}+15a_{2}^{2}+5a_{2}^{3})(Pr^2)^3\right.
\nonumber\\
&&\ \ +6(3+15a_{2}+15a_{2}^{2}+5a_{2}^{3})(Pr^2)^{2}(Qr^2)
\nonumber\\
&&\ \ +12(9+45a_{2}+65a_{2}^{2}+25a_{2}^{3})(Pr^2)(Qr^2)^2
\nonumber\\
&&\ \ +8(3+5a_{2})^3(Qr^2)^3
\nonumber\\
&&\ \ +15(21+27a_{2}+15a_{2}^{2}+5a_{2}^{3})(Pr^2)^2
\nonumber\\
&&\ \ +60(3+13a_{2}+5a_{2}^{2}-a_{2}^{3})(Pr^2)(Qr^2)
\nonumber\\
&&\ \ +60(9+39a_{2}+43a_{2}^{2}+5a_{2}^{3})(Qr^2)^2
\nonumber\\
&&\ \ +24(53-39a_{2}-142a_{2}^{2}-65a_{2}^{3})(Pr^2)
\nonumber\\
&&\ \ +48(9+33a_{2}-16a_{2}^{2}-15a_{2}^{3})(Qr^2)
\nonumber\\
&&\ \ \left.
+10(95-915a_{2}-1539a_{2}^{2}-633a_{2}^{3})\right]
\\
\zeta^{T}_{8}(0)&=&\frac{1}{75600(1+a_{2})^4}\times
\nonumber\\
&&\ \left[15(9+28a_{2}+42a_{2}^{2}+28a_{2}^{3}+7a_{2}^{4})(Pr^2)^4
\right.
\nonumber\\
&&\ \ +30(4+28a_{2}+42a_{2}^{2}+28a_{2}^{3}+7a_{2}^{4})
(Pr^2)^3(Qr^2)
\nonumber\\
&&\ \ +90(16+112a_{2}+238a_{2}^{2}+182a_{2}^{3}+49a_{2}^{4})
(Pr^2)^2(Qr^2)^2
\nonumber\\
&&\ \ +120(4+7a_{2})^2(4+14a_{2}+7a_{2}^{2})(Pr^2)(Qr^2)^3
\nonumber\\
&&\ \ +60(4+7a_{2})^4(Qr^2)^4
\nonumber\\
&&\ \ +140(36+90a_{2}+114a_{2}^{2}+76a_{2}^{3}+19a_{2}^{4})
(Pr^2)^3
\nonumber\\
&&\ \ +420(8+50a_{2}+42a_{2}^{2}+8a_{2}^{3}-a_{2}^{4})
(Pr^2)^2(Qr^2)
\nonumber\\
&&\ \ +840(32+200a_{2}+348a_{2}^{2}+167a_{2}^{3}+14a_{2}^{4})
(Pr^2)(Qr^2)^2
\nonumber\\
&&\ \ +560(4+7a_{2})^2(8+22a_{2}+5a_{2}^{2})(Qr^2)^3
\nonumber\\
&&\ \ +1680(38+54a_{2}+16a_{2}^{2}-3a_{2}^{3})(Pr^2)^2
\nonumber\\
&&\ \ +2520(12+66a_{2}-a_{2}^{2}-52a_{2}^{3}-18a_{2}^{4})
(Pr^2)(Qr^2)
\nonumber\\
&&\ \ +2520(48+264a_{2}+355a_{2}^{2}+34a_{2}^{3}-42a_{2}^{4})
(Qr^2)^2
\nonumber\\
&&\ \ +240(1270-992a_{2}-6515a_{2}^{2}-6318a_{2}^{3}
-1890a_{2}^{4})(Pr^2)
\nonumber\\
&&\ \ +720(120+570a_{2}-683a_{2}^{2}-1126a_{2}^{3}-336a_{2}^{4})
(Qr^2)
\nonumber\\
&&\ \ \left.
+56(6903-48948a_{2}-134742a_{2}^{2}-113908a_{2}^{3}
-32437a_{2}^{4})\right]
\\
\zeta^{T}_{10}(0)&=&\frac{1}{5443200(1+a_{2})^5}\times
\nonumber\\
&&\ \left[15(11+45a_{2}+90a_{2}^{2}+90a_{2}^{3}+45a_{2}^{4}
+9a_{2}^{5})(Pr^2)^5\right.
\nonumber\\
&&\ \ +30(5+45a_{2}+90a_{2}^{2}+90a_{2}^{3}+45a_{2}^{4}
+9a_{2}^{5})(Pr^2)^4(Qr^2)
\nonumber\\
&&\ \ +120(25+225a_{2}+630a_{2}^{2}+720a_{2}^{3}+387a_{2}^{4}
+81a_{2}^{5})(Pr^2)^3(Qr^2)^2
\nonumber\\
&&\ \ +240(125+1125a_{2}+3600a_{2}^{2}+5094a_{2}^{3}
+3159a_{2}^{4}+729a_{2}^{5})(Pr^2)^2(Qr^2)^3
\nonumber\\
&&\ \ +240(5+9a_{2})^3(5+18a_{2}+9a_{2}^{2})(Pr^2)(Qr^2)^4
\nonumber\\
&&\ \ +96(5+9a_{2})^5(Qr^2)^5
\nonumber\\
&&\ \ +45(275+995a_{2}+1830a_{2}^{2}+1830a_{2}^{3}
+915a_{2}^{4}+183a_{2}^{5})(Pr^2)^4
\nonumber\\
&&\ \ +360(25+205a_{2}+270a_{2}^{2}+150a_{2}^{3}+39a_{2}^{4}
+3a_{2}^{5})(Pr^2)^3(Qr^2)
\nonumber\\
&&\ \ +1080(125+1025a_{2}+2450a_{2}^{2}+2098a_{2}^{3}
+725a_{2}^{4}+81a_{2}^{5})(Pr^2)^2(Qr^2)^2
\nonumber\\
&&\ \ +1440(625+5125a_{2}+14700a_{2}^{2}+17820a_{2}^{3}
+8451a_{2}^{4}+1215a_{2}^{5})(Pr^2)(Qr^2)^3
\nonumber\\
&&\ \ +720(5+9a_{2})^3(25+70a_{2}+21a_{2}^{2})(Qr^2)^4
\nonumber\\
&&\ \ +60(5815+16805a_{2}+24400a_{2}^{2}+22290a_{2}^{3}
+11361a_{2}^{4}+2301a_{2}^{5})(Pr^2)^3
\nonumber\\
&&\ \ +360(545+4033a_{2}+1934a_{2}^{2}-2838a_{2}^{3}
-2427a_{2}^{4}-495a_{2}^{5})(Pr^2)^2(Qr^2)
\nonumber\\
&&\ \ +720(2725+20165a_{2}+39718a_{2}^{2}+19410a_{2}^{3}
-2679a_{2}^{4}-2403a_{2}^{5})(Pr^2)(Qr^2)^2
\nonumber\\
&&\ \ +480(13625+100825a_{2}+257210a_{2}^{2}+260406a_{2}^{3}
+80757a_{2}^{4}+729a_{2}^{5})(Qr^2)^3
\nonumber\\
&&\ \ +120(37589+63843a_{2}+3462a_{2}^{2}-53032a_{2}^{3}
-34947a_{2}^{4}-6183a_{2}^{5})(Pr^2)^2
\nonumber\\
&&\ \ +480(3805+25113a_{2}-15072a_{2}^{2}-63632a_{2}^{3}
-40959a_{2}^{4}-7875a_{2}^{5})(Pr^2)(Qr^2)
\nonumber\\
&&\ \ +480(19025+125565a_{2}+193326a_{2}^{2}-6850a_{2}^{3}
-103923a_{2}^{4}-33615a_{2}^{5})(Qr^2)^2
\nonumber\\
&&\ \ +4320(5920-2712a_{2}-42445a_{2}^{2}-65558a_{2}^{3}
-40680a_{2}^{4}-9300a_{2}^{5})(Pr^2)
\nonumber\\
&&\ \ +17280(350+2030a_{2}-4172a_{2}^{2}-10153a_{2}^{3}
-6280a_{2}^{4}-1250a_{2}^{5})(Qr^2)
\nonumber\\
&&\ \ +8(5745607-30218365a_{2}-125264810a_{2}^{2}
-163829090a_{2}^{3}-94569445a_{2}^{4}
\nonumber\\
&&\ \ \ \left.
-20749889a_{2}^{5})
\right]
\end{eqnarray}

\end{document}